\documentclass[aps,prb,reprint,showpacs,superscriptaddress,longbibliography]{revtex4-2}
\usepackage{mathtools,amssymb,graphicx,units,bm}
\usepackage{amsmath}
\usepackage[plainpages=false,pdfpagelabels,colorlinks=true,linkcolor=red,urlcolor=blue,citecolor=blue,pdftitle={Title},pdfauthor={},pdfdisplaydoctitle=true,pdfduplex=DuplexFlipLongEdge]{hyperref}
\usepackage{verbatim}
\usepackage[english]{babel}
\usepackage{color}

\usepackage[normalem]{ulem}

\begin{document}

\title{Superfluidity without charge order in the attractive Hubbard model on the kagome lattice}
\author{Xiaodong Jin}
\affiliation{Beijing Computational Science Research Center, Beijing 100193, China}
\author{Yingping Mou}
\affiliation{College of Physics and Electronic Engineering, Chongqing Normal University, Chongqing 401331, China}
\author{Rubem Mondaini}
\email{rmondaini@uh.edu}
\affiliation{Department of Physics, University of Houston, Houston, Texas 77204, USA}
\affiliation{Texas Center for Superconductivity, University of Houston, Houston, Texas 77204, USA}
 
\begin{abstract}
Using auxiliary-field quantum Monte Carlo simulations, we study the zero-temperature attractive Hubbard model on the kagome lattice at various relevant electronic fillings. The low-energy physics at $2/3$-density is influenced by the Dirac point at the Fermi energy, wherein we unveil a superfluid transition at a critical attractive interaction $U_c/t=-4.58(3)$, which belongs to the chiral-XY universality class. This U(1) symmetry-breaking is not accompanied by charge order, even for substantially large interaction strengths, in a regime where a description in terms of hardcore bosons becomes increasingly suitable. An investigation of the latter shows that charge ordering at $1/3$-filling only occurs at interaction strengths much larger than those corresponding to the mapping to the original fermionic model. Additionally, for densities exhibiting van Hove singularities in the non-interacting density of states, our results show that any attractive interaction gives rise to superfluidity, yet again without charge ordering, contrary to recent studies employing mean-field theory.
\end{abstract}

\maketitle

\section{Introduction}

The kagome lattice provides a minimal setting in which geometric frustration, sublattice interference, flat bands, Dirac points, and van Hove singularities combine to amplify electronic correlation effects and generate competing ordered states~\cite{Kiesel2012,Kiesel2013,Wang2013,Jiang2023}. This makes it a natural starting point for understanding the intertwined charge order, topological responses, and superconductivity observed in kagome metals such as AV$_3$Sb$_5$~\cite{Jiang2023,Wilson2024}. Yet the appropriate minimal description in these materials remains under debate.

For instance, weak-coupling and Ginzburg-Landau descriptions of electronically driven $3Q$ charge order~\cite{Denner2021,Lin2021}, extended-Hubbard approaches near van Hove filling~\cite{Wang2013,Ferrari2022}, chiral-flux or topological-charge density wave (CDW) scenarios~\cite{Feng2021,Lin2021}, and structural or electron-phonon mechanisms~\cite{Tan2021,Christensen2021,Xie2022} have all been proposed to account for the observed CDW, time-reversal-symmetry-breaking signatures~\cite{Mielke2022}, and superconducting phenomenology~\cite{Wu2021}. In particular, the extent to which phonon degrees of freedom are essential in this description, rather than secondary to electronically driven kagome instabilities, remains unclear~\cite{Li2021,Christensen2021,Xie2022,Ferrari2022,Bradley2023,Hu2024}.

Focusing on superconductivity, previous studies have argued that the kagome geometry can support multiple unconventional pairing channels, including chiral spin-singlet $d_{x^2-y^2}+id_{xy}$ pairing~\cite{Ko2009,Yu2012,Wang2013}, $d$-wave pairing tendencies that may condense into a chiral $d+id$ state~\cite{Wen2022}, and triplet $f$-wave pairing~\cite{Kiesel2013}. More recent weak-coupling studies motivated by AV$_3$Sb$_5$ further point to a near-degeneracy among singlet, triplet, and sublattice-modulated pairing channels~\cite{Wu2021,Romer2022,Schwemmer2024}. 

On the experimental side, several observations in AV$_3$Sb$_5$ are consistent with a conventional spin-singlet, nodeless superconducting component, including the suppression of the Knight shift and a Hebel-Slichter coherence peak below $T_c$~\cite{Mu2021}, nodeless penetration-depth behavior~\cite{Duan2021,Gupta2022}, and impurity responses compatible with sign-preserving pairing~\cite{Xu2021}. This picture is nevertheless incomplete, since thermal-transport measurements have been interpreted as evidence for residual low-energy excitations and possible nodal behavior~\cite{Zhao2024}, STM/STS reveals multigap spectra and deep gap minima~\cite{Xu2021}, and spatially modulated superconducting features have been interpreted as evidence for pair-density-wave order~\cite{Chen2021,Jiang2023}. Thus, while a local spin-singlet component appears experimentally relevant, superconductivity in AV$_3$Sb$_5$ is not captured by a featureless BCS description.

Motivated by the evidence for a local spin-singlet component, we study the attractive Hubbard model on the kagome lattice as a controlled limit of onsite, instantaneous pairing. While this model is unlikely to be the starting point for reproducing the full phenomenology of AV$_3$Sb$_5$, it offers an angle from which to explore how kagome band geometry, including Dirac points, van Hove singularities, and flat-band effects, reshapes the competition between superfluidity and charge order when the pairing interaction is purely local.

This question is particularly useful because attractive Hubbard models on bipartite lattices possess pseudospin constraints that tightly relate superconducting and charge-density-wave correlations~\cite{Micnas1990,Lee2009}. On the non-bipartite kagome lattice, by contrast, geometric frustration can suppress simple charge ordering and thereby favor superconductivity. A recent study combining mean-field theory and finite-temperature determinantal quantum Monte Carlo (QMC) suggested that the attractive kagome Hubbard model can nevertheless host charge-density-wave order at sufficiently large attraction~\cite{Zhu2023}. As we show below, using ground-state QMC simulations~\cite{Loh1992, Assaad2008}, this charge-ordering tendency is not supported at $T=0$.

\section{Model}

\begin{figure}[!tb] 
  \centering
  \includegraphics[width=\columnwidth]{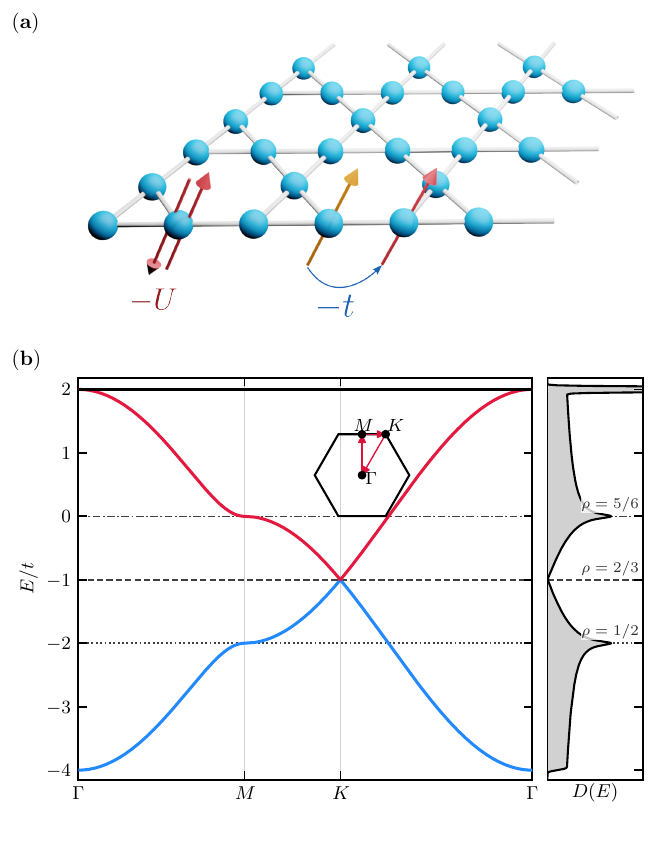}
  \caption{
    (a) Geometry of the kagome lattice with linear size $L=3$. The on-site attractive interaction $-U$, with $U>0$, and the nearest-neighbor hopping amplitude $-t$ are illustrated. (b) Noninteracting band structure along the high-symmetry path $\Gamma$-$M$-$K$-$\Gamma$ in the first Brillouin zone (inset), together with the density of states $D(E)$ (right). The dotted, dashed, and dash-dotted horizontal lines indicate the
    noninteracting Fermi energies at total densities $\rho=1/2$, $2/3$, and $5/6$, respectively.
    The densities $\rho=1/2$ and $5/6$ correspond to van Hove singularities, whereas $\rho=2/3$ is the Dirac filling.
  }
  \label{fig:kagome-lattice}
\end{figure}

We study the attractive Hubbard model on the kagome lattice,
\begin{equation}
\label{eq:Hamiltonian}
\hat{\mathcal H}
=
-t\sum_{\langle ij\rangle,\sigma}
\left(
\hat c^{\dagger}_{i\sigma}\hat c_{j\sigma}^{\phantom{\dagger}}
+\mathrm{H.c.} \right) -U\sum_i
\hat n_{i\uparrow}\hat n_{i\downarrow},
\end{equation}
where $\hat c_{i\sigma}$ ($\hat c^{\dagger}_{i\sigma}$) annihilates (creates) a fermion with spin $\sigma$ at lattice site $i$, and $\hat n_{i\sigma}=\hat c^{\dagger}_{i\sigma}\hat c_{i\sigma}$ is the corresponding number operator. Here, $t>0$ is the nearest-neighbor hopping amplitude and $U>0$ denotes the magnitude of the on-site attraction, as illustrated in Fig.~\ref{fig:kagome-lattice}(a). 

We consider kagome lattices with linear size $L$ and $N=3L^2$ sites. For the total electronic density, $\rho = \frac{1}{N} \sum_{i,\sigma} \left\langle \hat n_{i\sigma}\right\rangle$, we focus on $\rho=1/2$ and $5/6$, which coincide with van Hove singularities of the noninteracting ($U=0$) density of states, and on $\rho=2/3$, corresponding to the Dirac filling [Fig.~\ref{fig:kagome-lattice}(b)]. We employ ground-state projector auxiliary-field quantum Monte Carlo simulations~\cite{Loh1992, Assaad2008} in the canonical ensemble. For the spin-balanced ($N_\uparrow = N_\downarrow$) attractive model considered here, the simulations are free of the fermion sign problem~\cite{Loh1990, Mondaini2022}. The imaginary-time evolution is discretized using a time step $\Delta\tau t=0.1$, and the projection length $\Theta t=70$ is chosen sufficiently large that the measured observables are converged within statistical uncertainties and correctly capture the ground-state physics.

For nearest-neighbor hopping, translational invariance allows the noninteracting Hamiltonian to be written in momentum space as 
\begin{equation}
\hat{\mathcal H}_0 =
\sum_{\mathbf{k},\sigma}
\hat{\Psi}_{\mathbf{k}\sigma}^{\dagger}
\mathcal{H}_0(\mathbf{k})
\hat{\Psi}_{\mathbf{k}\sigma},
\end{equation}
where the spinor $ \hat{\Psi}_{\mathbf{k}\sigma} =
\begin{pmatrix}
\hat c_{\mathbf{k}1\sigma} &
\hat c_{\mathbf{k}2\sigma} &
\hat c_{\mathbf{k}3\sigma}
\end{pmatrix}^{T}$ is defined in the three-sublattice basis of the kagome unit cell. The corresponding Bloch Hamiltonian is
\begin{equation}
\label{eq:bloch-hamiltonian}
\mathcal{H}_0(\mathbf{k})
=
-2t
\begin{pmatrix}
0              & \cos k_1 & \cos k_2 \\
\cos k_1       & 0        & \cos k_3 \\
\cos k_2       & \cos k_3 & 0
\end{pmatrix},
\end{equation}
where $k_n=\mathbf{k}\cdot\mathbf{\delta}_n$, with $
\mathbf{\delta}_1=(1,0)$ , $\mathbf{\delta}_2=\frac{1}{2}(-1,\sqrt{3})$, and $\mathbf{\delta}_3=\frac{1}{2}(-1,-\sqrt{3})$, denoting the three nearest-neighbor bond directions~\cite{Guo2009}.

Diagonalizing Eq.~\eqref{eq:bloch-hamiltonian} yields an exactly flat
band, $\varepsilon_{0}(\mathbf{k})=2t$, and two dispersive bands, $
\varepsilon_{\pm}(\mathbf{k}) = t\left[-1\pm\sqrt{4f(\mathbf{k})-3}
\right]$, where $f(\mathbf{k}) = \cos^{2}k_1+\cos^{2}k_2+\cos^{2}k_3$. Each band is twofold degenerate owing to the spin degree of freedom. The two dispersive bands meet linearly at the inequivalent Brillouin-zone corners $\mathbf{K}_{\pm}=(\pm2\pi/3,0)$, forming Dirac cones at energy $\varepsilon_{\pm}(\mathbf{K}_{\pm})=-t$. Consequently, filling the lowest band corresponds to the total density $\rho=2/3$. The upper dispersive band touches the flat band quadratically at $\Gamma$, where both have energy $2t$. Finally, the saddle points at $M$ occur at energies $-2t$ and $0$, producing the two van Hove
singularities associated with the densities $\rho=1/2$ and $5/6$, respectively, as shown in Fig.~\ref{fig:kagome-lattice}(b).

\section{RESULTS}

Our goal is to determine the ordered phases and associated quantum phase transitions at the three fillings highlighted above. We begin with $\rho=2/3$, where the Fermi energy crosses the Dirac points and the vanishing density of states stabilizes the semimetal against weak short-range interactions. Interaction-driven transitions of two-dimensional Dirac fermions~\cite{Paiva2005, Meng2010, Sorella2012} depend on both the symmetry of the order parameter and whether the corresponding critical fluctuations couple to gapless fermions. For example, in the repulsive honeycomb and $\pi$-flux Hubbard models, the transition into an antiferromagnetic state with increasing interactions is described by the chiral-Heisenberg Gross-Neveu universality class~\cite{Assaad2013,Toldin2015,Otsuka2016,Otsuka2020}. In contrast, in the Kane-Mele-Hubbard model the fermions remain gapped across the onset of transverse antiferromagnetic order, and the transition follows the more conventional 3D-XY universality class~\cite{Hohenadler2012}.

For attractive interactions, the half-filled honeycomb model provides a distinct reference point because superconducting and charge-density-wave orders are related by an enlarged pseudospin symmetry~\cite{Lee2009}. Indeed, under a particle-hole transformation of one spin component, $\hat c_{i,\downarrow}\rightarrow(-1)^i\hat c_{i,\downarrow}^{\dagger}$, the longitudinal spin operator of the repulsive model, $\hat m_i^z \equiv \hat n_{i\uparrow} - \hat n_{i\downarrow}$, transforms as $\hat m_i^z\rightarrow \hat n_i-1$, such that antiferromagnetic correlations in this channel map onto sublattice-staggered charge correlations. Likewise, the transverse-spin correlations map onto onsite pairing correlations, $\langle \hat m_i^+\hat m_j^-\rangle \rightarrow (-1)^{i+j} \langle \hat\Delta_i^\dagger\hat\Delta_j\rangle$, where $\hat\Delta_i=\hat c_{i,\downarrow}\hat c_{i,\uparrow}$ is the pair annihilation operator at site $i$. That is, the staggered transverse order of the repulsive model therefore becomes uniform $s$-wave superconductivity in the attractive model. Consequently, the CDW order parameter and the real and imaginary components of the pairing field form a three-component pseudospin vector and exhibit degenerate correlations at half-filling. 

Nonetheless, this correspondence relies on the bipartite structure of the honeycomb lattice and does not extend to the nonbipartite kagome lattice. A closer comparison is the attractive $\pi$-flux triangular-lattice Hubbard model, where a finite attraction drives a Dirac semimetal into a zero-momentum $s$-wave superconductor without accompanying CDW order~\cite{Otsuka2018}, where the corresponding transition was identified with the chiral-XY Gross-Neveu universality class instead. 


\subsection{Finite-size scaling of the superfluid transition}
\begin{figure}[!tb] 
    \includegraphics[width=1\columnwidth]{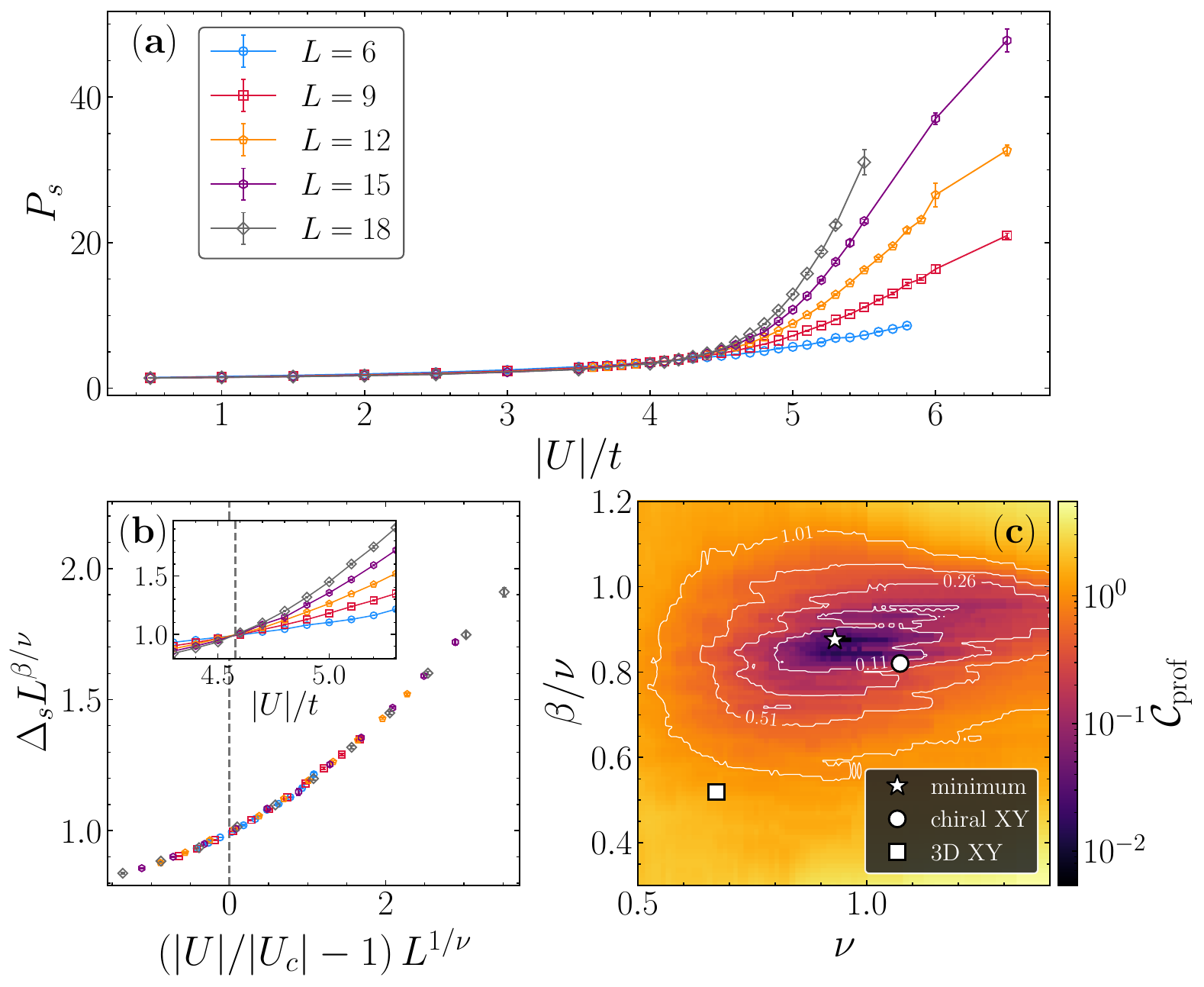}   
    \caption{Finite-size scaling of the superconducting pair structure factor. (a) Pair structure factor $P_s$ as a function of $|U|/t$ for the indicated linear system sizes $L$. (b) Collapse of the scaled order parameter $\Delta_sL^{\beta/\nu}$ as a function of $(|U|/|U_c|-1)L^{1/\nu}$, using the parameters obtained by minimizing the collapse cost. The inset shows the same scaled observable as a function of $|U|/t$ near the critical region; the dashed line denotes the optimal $|U_c|/t$. (c) Profiled collapse cost $\mathcal{C}_{\mathrm{prof}}(\nu,\beta/\nu) =\min_{|U_c|/t}\mathcal{C}(|U_c|/t,\nu,\beta/\nu)$, calculated using $4.3\leq |U|/t\leq5.3$. The star marks the global minimum at $|U_c|/t=4.58(3)$, $\nu=0.93(3)$, and $\beta/\nu=0.88(1)$, corresponding to $\beta=0.82(3)$. The open circle and square indicate the chiral-XY and 3D-XY benchmark exponents, respectively~\cite{Otsuka2018,Campostrini2006}.}
    \label{fig:scaling-ps} 
\end{figure}
We begin by characterizing the onset of superfluid order through the uniform onsite $s$-wave pair structure factor
\begin{equation}
    P_s(L,U)=\frac{1}{N}\sum_{i,j}
    \langle
    \hat{\Delta}_{i}^\dagger
    \hat{\Delta}_{j}^{\phantom{\dagger}}
    \rangle\ ;
\label{eq:Ps}
\end{equation}
long-range pair order implies $P_s$ is extensive in the system size in the thermodynamic limit. Indeed, as shown in Fig.~\ref{fig:scaling-ps}(a), $P_s$ develops an increasingly pronounced system-size dependence for sufficiently strong attraction, consistent with the onset of long-range superconducting order. We therefore define the finite-size estimator of the order parameter as $\Delta_s(L,U)=\sqrt{P_s(L,U)/N}$~\cite{Otsuka2018}. Near a continuous quantum critical point, the order parameter obeys the finite-size scaling form
\begin{equation}
    \Delta_s(L,U)
    =L^{-\beta/\nu}
    f_s(uL^{1/\nu}),
\label{eq:scaling_relation}
\end{equation}
where $u=\frac{|U|}{|U_c|}-1$ is the reduced coupling,  $\beta$ and $\nu$ are the order-parameter and correlation-length critical exponents, and $f_s$ is a universal scaling function. At criticality, Eq.~\eqref{eq:scaling_relation} gives $\Delta_s(L,U_c)\propto L^{-\beta/\nu}$, such that $\Delta_sL^{\beta/\nu}$ is independent of $L$ to leading order. For extracting the critical coupling and the universality class, we simultaneously determine the three parameters $|U_c|/t$, $\nu$, and $\beta/\nu$ from the quality of the data collapse.

To quantify it, for each trial triplet $(|U_c|/t,\nu,r)$, with $r=\beta/\nu$, we define $x_j=(|U_j|/|U_c|-1)L_j^{1/\nu}$ and $y_j=\Delta_s(L_j,U_j)L_j^r$. After ordering the observations by increasing $x_j$, we calculate a cost function $\mathcal{C}=[\sum_j|y_{j+1}-y_j|]/[\max_j(y_j)-\min_j(y_j)]-1$~\cite{Suntajs2020,Aramthottil2021,Mondaini2023}. Here, $U_c$ fixes the origin of the scaling coordinate, $\nu$ controls the horizontal size rescaling, and $\beta/\nu$ controls the vertical rescaling. Lower values of $\mathcal{C}$ indicate a smoother collapse, with $\mathcal{C}=0$ for a monotonic ordered sequence. Within that approach, we evaluate $\mathcal{C}$ over a three-dimensional grid in $(|U_c|/t,\nu,\beta/\nu)$. 

Figure~\ref{fig:scaling-ps}(c) shows the profiled cost,
$\mathcal{C}_{\mathrm{prof}}(\nu,\beta/\nu)
=\min_{|U_c|/t}\mathcal{C}$, such that each point represents the minimum cost obtained after optimizing over $U_c$. Using the coupling window $4.3\leq |U|/t\leq 5.3$ and all available system sizes, we obtain $|U_c|/t=4.58(3)$, $\nu=0.93(3)$, and $\beta/\nu=0.88(1)$, corresponding to $\beta=0.82(3)$. The quoted uncertainties are bootstrap standard deviations and are therefore conditional on the chosen coupling window and lattice sizes. These exponents are substantially closer to numerical estimates for the chiral-XY universality class in related Dirac-fermion models~\cite{Otsuka2018,Li2017} than to the high-precision 3D-XY values $\nu=0.6717(1)$ and $\beta=0.3486(1)$~\cite{Campostrini2006}, consistent with the gapless semimetallic character of the weak-coupling phase. The slightly smaller value of $\nu$ relative to existing chiral-XY estimates may reflect residual finite-size or scaling-window corrections. By contrast, 3D-XY criticality is expected when the fermionic single-particle excitations are already gapped and the transition is governed primarily by the establishment of phase coherence among preformed pairs, as occurs in bosonic or fermionic insulator-to-superconductor transitions~\cite{Mondaini2015,Xiaodong2022}.

\subsection{Single-particle excitations}

Further insight into the nature of the transition and its type can be inferred from the spectrum of excitations at this $\rho = 2/3$ filling. This can be quantified by the spectral function $A({\bf k},\omega)$, which can be extracted from the corresponding imaginary-time Green's functions:
\begin{equation}
    G({\bf k},\tau) =  \int d{\omega}\  \frac{e^{-\omega \tau}}{1+e^{-\omega/T}} A({\bf k},\omega)\ .
    \label{eq:Green function}
\end{equation}
Inverting this integral equation requires an analytical continuation, for which we use the Maximum Entropy approach~\cite{Assaad2022}.

\begin{figure}[!tb] 
  \includegraphics[width=1\columnwidth]{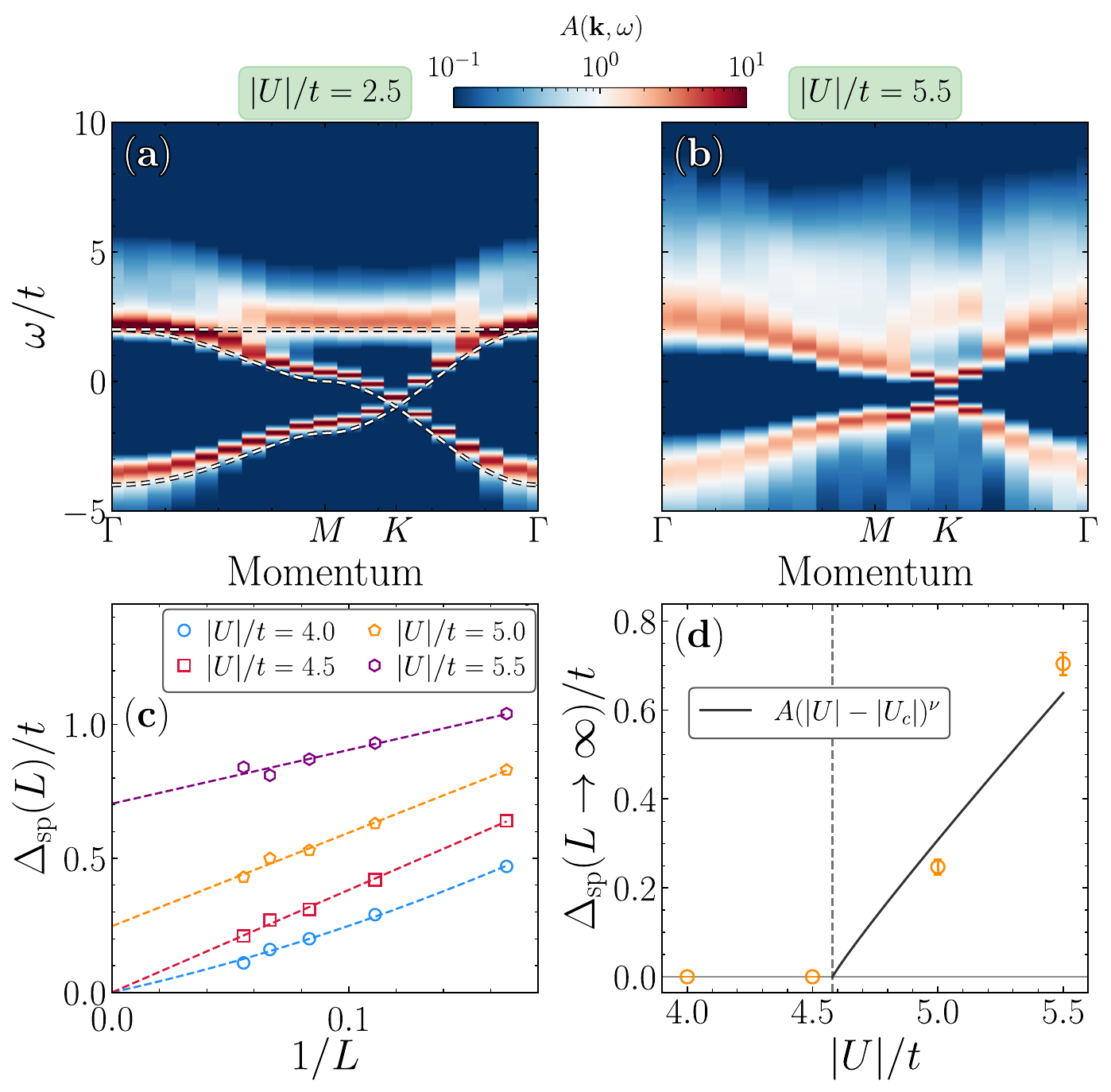}
  \caption{Single-particle spectral function and excitation gap across the superfluid transition. (a,b) Momentum-resolved spectral function $A(\mathbf{k},\omega)$ along the $\Gamma$–$M$–$K$–$\Gamma$ path for $L=18$ at the indicated interaction
  strengths. The dashed curves in (a) show the noninteracting kagome-lattice dispersion. (c) Polynomial fittings of the single-particle energy gap at the $K$ Dirac point with the inverse of the linear lattice size from $L=6$ to $L=15$ are performed for different values of $U$ at the  -- except for $|U|/t=4$ where we use a third-order polynomial with zero constant term (d) Extrapolated gap $\Delta_{\mathrm{sp}}(L\to\infty)$ versus $|U|/t$. The solid curve is a fit to $A(|U|-|U_c|)^{z\nu}$, with $|U_c|/t=4.58$, $\nu=0.93$ and $z=1$ fixed from the finite-size scaling analysis of Fig.~\ref{fig:scaling-ps}; the vertical dashed line marks $|U_c|/t$.}
  \label{fig:single-gap}
\end{figure}

Figures~\ref{fig:single-gap}(a) and \ref{fig:single-gap}(b) show the spectral function along the high-symmetry path $\Gamma$-$M$-$K$-$\Gamma$ for representative interaction strengths below and above the critical coupling, respectively, in an $L=18$ lattice. On the weak-coupling side, the interacting spectrum closely follows the noninteracting dispersion [Fig.~\ref{fig:kagome-lattice}(b)]: apart from a moderate band renormalization, the Dirac crossing at $K$ remains clearly visible. With increasing $|U|$, spectral weight is progressively removed from $\omega=0$, and a single-particle gap opens at the Dirac point upon entering the superfluid phase. 

To quantify this behavior, we extract a finite-size peak-to-peak single-particle gap from the analytically continued spectra at the $K$ point. For each pair $(U,L)$, we identify the dominant peaks of the particle-addition and particle-removal spectra, denoted by $\omega^{\rm pk}_+(K)$ and $\omega^{\rm pk}_-(K)$, respectively. The single-particle gap is therefore $\Delta_{\rm sp}(L)=\omega^{\rm pk}_+(K)-\omega^{\rm pk}_-(K)$.
Figure~\ref{fig:single-gap}(c) compiles the finite-size results, showing that in the semimetallic phase, $\Delta_{\rm sp}(L)$ vanishes upon approaching the thermodynamic limit, whereas a finite extrapolated gap emerges for $|U|>|U_c|$. Near a continuous quantum critical point, the correlation length diverges as $\xi\sim\bigl||U|-|U_c|\bigr|^{-\nu}$, while the characteristic excitation energy scales as $\xi^{-z}$~\cite{Sachdev2011}. The single-particle gap should therefore open according to $\Delta_{\rm sp}=A(|U|-|U_c|)^{z\nu}$. As shown in Fig.~\ref{fig:single-gap}(d), the extrapolated gaps are consistent with this scaling form when $|U_c|/t=4.58$ and $\nu=0.93$ are fixed from the order-parameter analysis and the relativistic value $z=1$ is assumed. Thus, the simultaneous emergence of superfluid order and a fermionic excitation gap is consistent with a continuous semimetal-to-superfluid transition governed by chiral-XY criticality. Other local quantities can also be used as local proxies for the transition, see Appendix.

\subsection{Absence of the CDW order}
From the previous discussion, we find that increasing the attractive interaction breaks the U(1) symmetry and leads to the emergence of a superfluid phase transition. In kagome materials such as AV$_3$Sb$_5$ (A = Cs, K, Rb), the vanadium atoms form stacked ideal kagome layers, giving rise to a variety of correlated electronic phases, including charge-density-wave (CDW) order observed below $T_{\rm CDW} \approx 80$–$110$ K~\cite{Ilija2021}. Since mean-field calculations have also predicted the formation of CDW states satisfying the triangle rule in this very same model~\cite{Zhu2023}, we examine whether CDW order appears in our unbiased calculations.

\begin{figure}[!tb]
  \includegraphics[width=1\columnwidth]{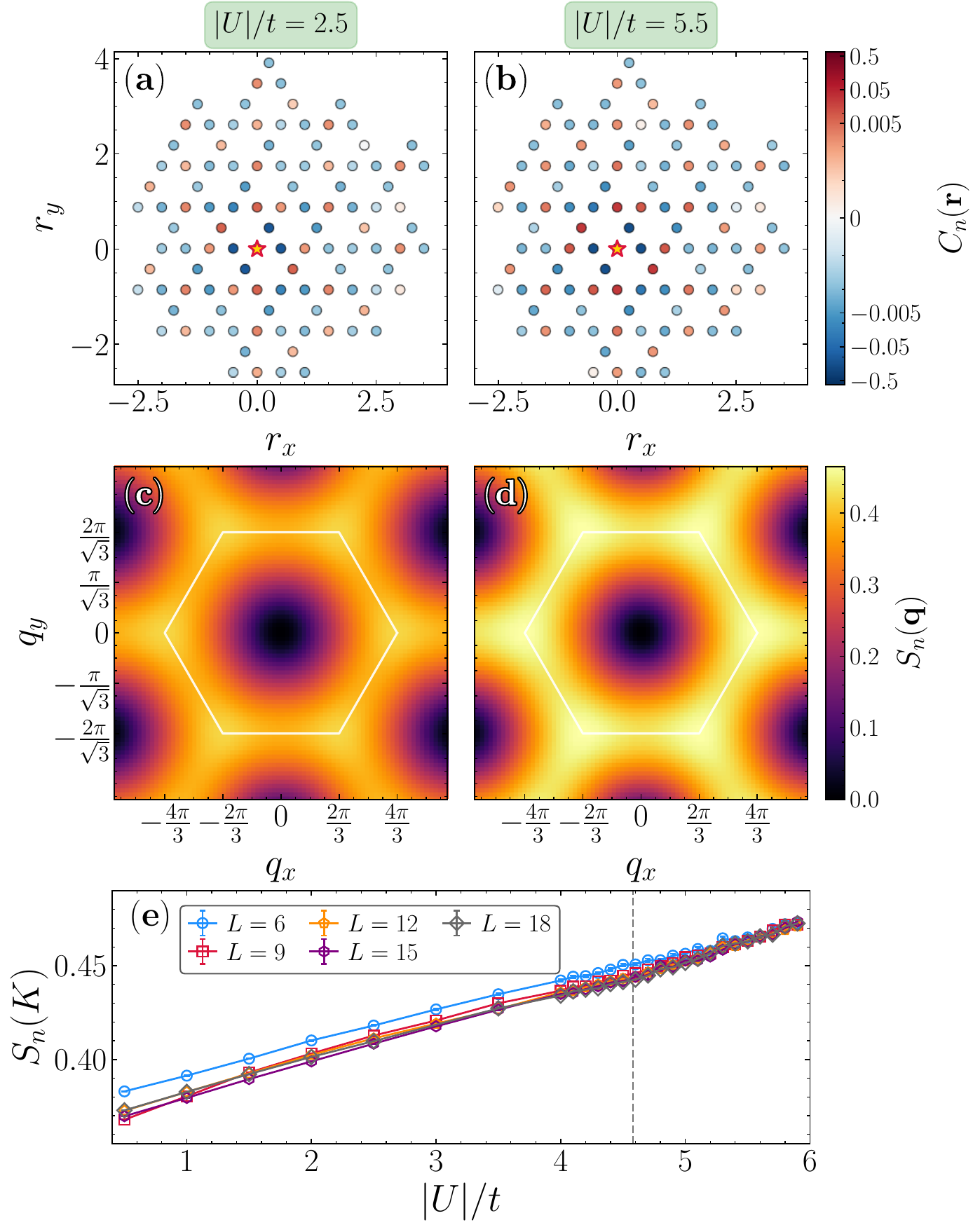}
    \caption{Real- and momentum-space density correlations across the superfluid transition. (a,b) Connected density correlation function $C_n(\mathbf{r})=\langle\hat n_{\mathbf{0}}\hat n_{\mathbf{r}}\rangle-\langle\hat n_{\mathbf{0}}\rangle\langle\hat n_{\mathbf{r}}\rangle$ for $L=6$ at the indicated interaction strengths; the star marks the reference site. (c,d)   Corresponding unit-cell density structure factor $S_n(\mathbf{q})$, obtained by Fourier transforming the real-space correlations and summing over the sublattice indices. The white hexagon delineates the first Brillouin zone. (e) Density structure factor $S_n(K)$ at the Brillouin-zone corner as a function of $|U|/t$ for the indicated system sizes; the vertical dashed line marks the critical interaction $|U_c|/t=4.58$ obtained from Fig.~\ref{fig:scaling-ps}.}
  \label{fig:Scdw}
\end{figure}

To investigate possible charge ordering, we first examine the connected real-space density correlation function $C_n(\mathbf r)=\langle\delta\hat n_{\mathbf 0}\delta\hat n_{\mathbf r}\rangle=\langle\hat n_{\mathbf 0}\hat n_{\mathbf r}\rangle-\langle\hat n_{\mathbf 0}\rangle\langle\hat n_{\mathbf r}\rangle$, where $\delta\hat n_{\mathbf r}=\hat n_{\mathbf r}-\langle\hat n_{\mathbf r}\rangle$ and $\mathbf 0$ denotes the reference site marked by the star in Figs.~\ref{fig:Scdw}(a) and \ref{fig:Scdw}(b), for $L=6$, $\rho=2/3$, and $|U|/t=2.5$ and $5.5$, respectively. Here, increasing $|U|$ enhances the short-distance correlations, but their rapid spatial decay persists, suggesting that the charge correlations remain short-ranged. To characterize their momentum dependence, we introduce the total density fluctuation in unit cell $\mathbf R$ by summing over the sublattice $\alpha$, $\delta\hat n_{\mathbf R}=\sum_\alpha\delta\hat n_{\mathbf R,\alpha}$, and calculate the unit-cell density structure factor
\begin{align}
  S_n(\mathbf q)
  &=\frac{1}{3N_c}\sum_{\mathbf R,\mathbf R'}e^{i\mathbf q\cdot(\mathbf R-\mathbf R')}\langle\delta\hat n_{\mathbf R}\delta\hat n_{\mathbf R'}\rangle \ ,
\end{align}
where $N_c = N/3$ is the number of unit cells. The resulting $S_n(\mathbf q)$ is shown in Figs.~\ref{fig:Scdw}(c) and \ref{fig:Scdw}(d), with the white hexagon denoting the first Brillouin zone. The maxima occur at the $K$ points and increase moderately with $|U|$, while remaining broad. Figure~\ref{fig:Scdw}(e) compiles this structure factor for different system sizes as one varies the interaction strength. Although $S_n(K)$ increases overall with attraction, it exhibits little systematic growth with system size. Since long-range CDW order would produce an extensive $S_n(K)$, this weak size dependence indeed supports the absence of long-range charge order for the range of interactions investigated.

\begin{figure}[!tb] 
  \includegraphics[width=1\columnwidth]{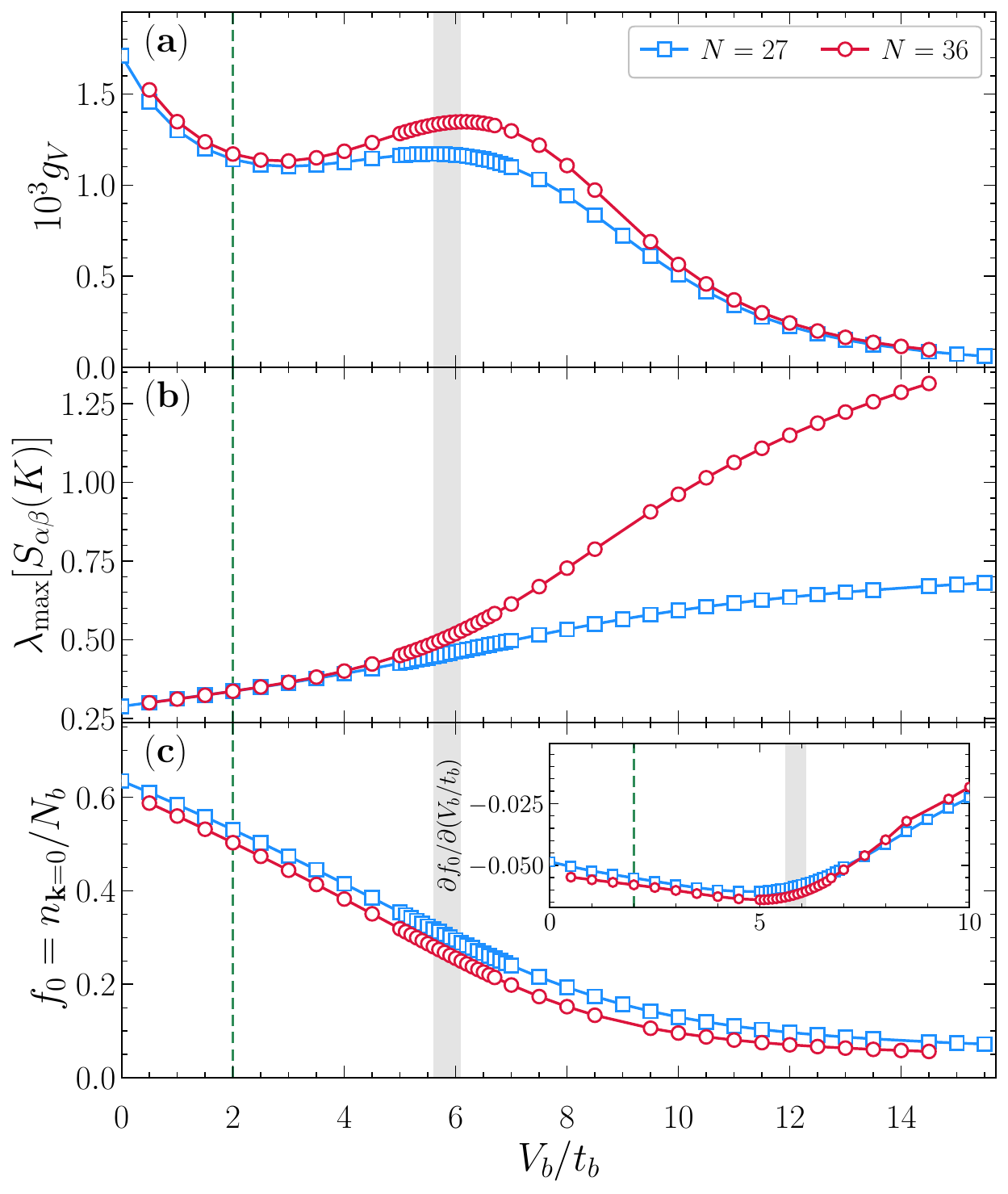}
  \caption{Exact-diagonalization results for the effective hard-core-boson model at filling $\rho_b=1/3$. (a) Fidelity susceptibility per site $g_V$, (b) largest eigenvalue $\lambda_{\max}[S_{\alpha\beta}(K)]$ of the sublattice-resolved connected density structure-factor matrix at the Dirac point $K=(2\pi/3,0)$, and (c) zero-momentum condensate fraction $f_0=n_{\mathbf{k}=0}/N_b$ as functions of $V_b/t_b$ for the indicated cluster sizes. The inset in (c) shows $\partial f_0/\partial(V_b/t_b)$. The gray shaded region spans the finite-size pseudocritical couplings identified by the fidelity maxima, $V_b/t_b=5.6$ and $6.1$ for $N=27$ and $36$, respectively, while the green dashed line marks the strong-coupling value $V_b/t_b=2$ obtained from the attractive-Hubbard
  mapping.}

  \label{fig:ED_result}
\end{figure}

\subsection{The strong interaction limit}
To test whether charge order could emerge at attractions stronger than those directly accessible in our fermionic calculations, we consider the strong-coupling regime $|U|\gg t$. Projecting onto the low-energy subspace of empty and doubly occupied sites and performing degenerate perturbation theory to second order in $t/|U|$ yields an effective hard-core-boson model~\cite{Micnas1990}. We define
$\hat b_i^\dagger=\hat c_{i\uparrow}^\dagger\hat c_{i\downarrow}^\dagger$
and $\hat n_{b,i}=\hat b_i^\dagger\hat b_i^{\phantom{\dagger}}$, such that the fermionic density $\rho=2/3$ corresponds to a bosonic density $\rho_b=1/3$. Up to an additive constant and a chemical-potential term, the effective Hamiltonian is
\begin{equation}
    \hat H_b
    =
    -t_b\sum_{\langle ij\rangle}
    \left(
    \hat b_i^\dagger\hat b_j^{\phantom{\dagger}}
    +\mathrm{H.c.}
    \right)
    +V\sum_{\langle ij\rangle}
    \hat n_{b,i}\hat n_{b,j},
    \label{eq:hard_core_bosons}
\end{equation}
where, for uniform nearest-neighbor hopping,
\begin{equation}
    t_b=\frac{2t^2}{|U|},
    \qquad
    V=\frac{4t^2}{|U|}=2t_b\ .
\end{equation}
Here, higher-order corrections are suppressed by additional powers of $t/|U|$. We subsequently treat $t_b$ and $V$ as independent parameters to locate the onset of charge order in the generalized hard-core-boson model. Comparing the resulting critical ratio $(V/t_b)_c$ with the physical strong-coupling value $V/t_b=2$ then determines whether the attractive Hubbard model approaches a charge-ordered state as $|U|/t\rightarrow\infty$.

We obtain the ground state using Krylov-subspace diagonalization~\cite{Balay,Slepc} on periodic kagome clusters containing $N=27$ and $36$ sites, whose geometries preserve the full spatial symmetry of the kagome lattice~\cite{Lauchli2011,Seman2015}. A finite-size precursor of the transition is identified through the fidelity susceptibility~\cite{Zanardi2006,CamposVenuti2007,Zanardi2007,You2007},
\begin{equation}
    g_V
    =
    \frac{2}{N}
    \frac{
    1-
    \left|
    \langle\Psi_0(t_b,V)
    \vert
    \Psi_0(t_b,V+\delta V)\rangle
    \right|
    }{(\delta V)^2},
\end{equation}
which quantifies the sensitivity of the ground-state wavefunction $|\Psi_0\rangle$ to a change in the interaction strength; we use $\delta V=10^{-3}t_b$. A peak that sharpens and grows with increasing system size provides evidence for a quantum phase transition and an estimate of its location~\cite{Mondaini2015,Xiaodong2022,Yang2007,Jia2011,Yi2021}.

Figure~\ref{fig:ED_result}(a) shows the fidelity susceptibility per site $g_V$; the broad local maxima occur at the finite-size pseudocritical couplings $V_b/t_b\simeq5.6$ and $6.1$ for $N=27$ and $36$, respectively. Their upward drift with system size is consistent with previous stochastic-series-expansion results, which located the superfluid-to-valence-bond-solid transition near
$(V_b/t_b)_c\simeq7.80$ at the tip of the $\rho_b=1/3$ insulating lobe~\cite{Isakov2006}. Indeed, to characterize density ordering while retaining the three-sublattice structure of the kagome unit cell, we define the connected correlation matrix $C_{\alpha\beta}(\mathbf{R})=\langle\delta\hat n_{0,\alpha}\delta\hat n_{\mathbf{R},\beta}\rangle$, where $\delta\hat n_{\mathbf{R},\alpha}=\hat n_{\mathbf{R},\alpha}-\langle\hat n_{\mathbf{R},\alpha}\rangle$. Its Fourier transform is $S_{\alpha\beta}(\mathbf{q})=\sum_{\mathbf{R}}e^{i\mathbf{q}\cdot\mathbf{R}}C_{\alpha\beta}(\mathbf{R})$. Figure~\ref{fig:ED_result}(b) shows the largest eigenvalue $\lambda_{\max}[S_{\alpha\beta}(K)]$. Unlike the scalar structure factor used in the PQMC analysis, this quantity retains the sublattice information and is therefore sensitive to density modulations within the kagome unit cell. Its increasing system-size dependence beyond the pseudocritical region signals the development of charge order. In the strong-coupling mapping of the attractive Hubbard model, the effective boson hopping and nearest-neighbor repulsion satisfy $t_b=2t^2/|U|$ and $V_b=4t^2/|U|$, fixing the physical ratio at $V_b/t_b=2$, as marked by the green dashed line. Because this value lies well inside the superfluid regime of the effective model, the asymptotic strong-attraction limit does not exhibit charge order. Together with the finite-$|U|$ PQMC results, this supports the absence of an intervening charge-ordered phase in the attractive Hubbard model.

Finally, Fig.~\ref{fig:ED_result}(c) shows the condensate fraction $f_0=n_{\mathbf{k}=0}/N_b$, where $n_{\mathbf{k}=0}=N^{-1}\sum_{i,j}\langle\hat b_i^\dagger\hat b_j^{\phantom{\dagger}}\rangle$. In the paired low-energy subspace, $\hat b_i^\dagger\leftrightarrow\hat\Delta_i^\dagger=\hat c_{i\uparrow}^\dagger\hat c_{i\downarrow}^\dagger$, so that $n_{\mathbf{k}=0}$ maps directly onto the fermionic pair structure factor $P_s=N^{-1}\sum_{i,j}\langle\hat\Delta_i^\dagger\hat\Delta_j\rangle$. Consequently, the plotted quantity corresponds to the pair structure factor normalized by the number of pairs, $f_0=P_s/N_b$. It remains finite in a superfluid and is suppressed upon entering the density-ordered insulating phase. The inset shows $\partial f_0/ \partial(V_b/t_b)$; its broad minimum provides a complementary finite-size signature of the loss of superfluid coherence.

\begin{figure}[!tb] 
  \includegraphics[width=1\columnwidth]{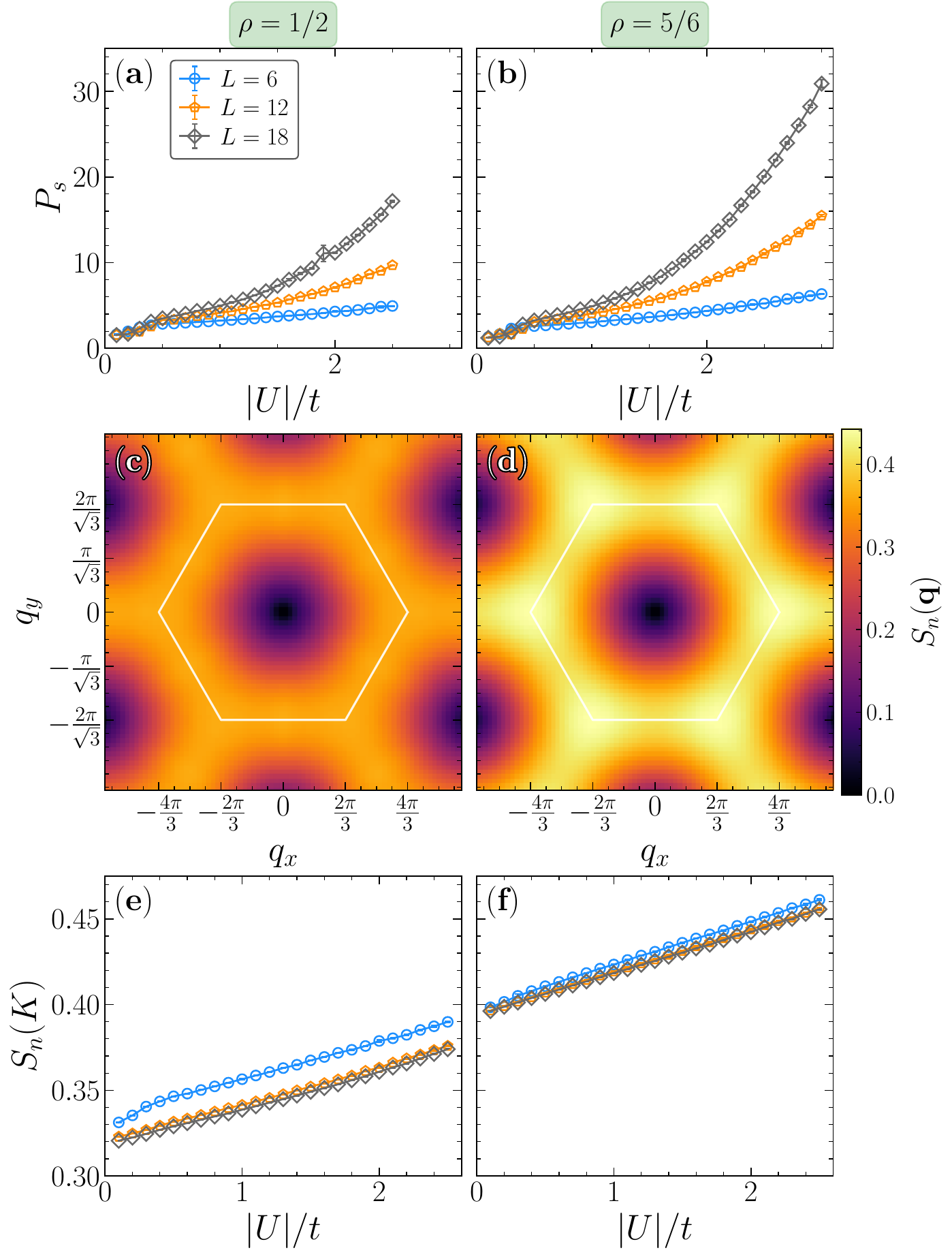}
\caption{Pairing and charge correlations at the two van Hove fillings. (a,b) Pair structure factor $P_s$ as a function of $|U|/t$ at (a) $\rho=1/2$ and (b) $\rho=5/6$ for the indicated lattice sizes. (c,d) Connected charge structure factor $S_n(\mathbf q)$ for $L=18$ and $|U|/t=2$ at (c) $\rho=1/2$ and (d) $\rho=5/6$; the white hexagon marks the first Brillouin zone. (e,f) Corresponding peak value $S_n(K)$ as a function of $|U|/t$. The strong size dependence of $P_s$, contrasted with the weak size dependence of $S_n(K)$, is consistent with superconducting order and the absence of long-range CDW order at both fillings.}
  \label{fig:van-hove-fillings}
\end{figure}

\subsection{Van Hove fillings}
We next consider the fillings $\rho=1/2$ and $\rho=5/6$, which correspond to the lower and upper van Hove singularities of the noninteracting kagome bands, respectively. At these fillings, the Fermi surface crosses the saddle points at the $M$-point [see Fig.~\ref{fig:kagome-lattice}(b)], producing a logarithmically divergent density of states. This situation contrasts with the Dirac point at $\rho=2/3$, where the density of states vanishes. For an on-site attraction, the enhanced Cooper susceptibility is expected to render the normal state unstable toward zero-momentum $s$-wave pairing for arbitrarily weak $|U|>0$ in the thermodynamic limit; hence, no finite critical attraction is expected.

Figures~\ref{fig:van-hove-fillings}(a) and \ref{fig:van-hove-fillings}(b) show the pair structure factor $P_s$ at the two van Hove fillings. In both cases, $P_s$ increases with $|U|/t$, and its growth with system size becomes increasingly pronounced as the attraction is strengthened, consistent with the development of superconducting correlations. The size dependence is particularly strong at $\rho=5/6$. Although these results are consistent with the weak-coupling expectation $|U_c|=0$, they do not establish it directly. On a finite cluster, the associated superconducting coherence length can exceed the accessible system sizes. Consequently, a clearly resolved growth of $P_s$ with $L$ emerges only once the attraction becomes sufficiently large.

Figures~\ref{fig:van-hove-fillings}(c) and \ref{fig:van-hove-fillings}(d) show the connected charge structure factor $S_n(\mathbf q)$ for $L=18$ and $|U|/t=2$. At both fillings, $S_n(\mathbf q)$ exhibits broad maxima at the $K$ points. To determine whether these correlations develop into long-range order, Figs.~\ref{fig:van-hove-fillings}(e) and \ref{fig:van-hove-fillings}(f) show the corresponding peak value $S_n(K)$ for different system sizes. Although $S_n(K)$ increases smoothly with attraction and is systematically larger at $\rho=5/6$ than at $\rho=1/2$, it does not grow with system size and instead exhibits a slight downward drift. This behavior is consistent with short-ranged charge correlations rather than a developing Bragg peak. Thus, over the interaction range investigated, we find no evidence of long-range CDW order at either van Hove filling, while the pronounced growth of $P_s$ identifies superconductivity as the leading interaction-driven instability.

\section{Summary and discussion}

We have investigated the ground-state physics of the attractive Hubbard model on the kagome lattice using unbiased auxiliary-field quantum Monte Carlo simulations. At the Dirac filling $\rho=2/3$, the vanishing noninteracting density of states stabilizes the semimetal up to a finite attraction $|U_c|/t=4.58(3)$. The critical scaling and the concomitant opening of the single-particle gap are consistent with a semimetal-to-superfluid transition in the chiral-XY universality class. At the two van Hove fillings, $\rho=1/2$ and $5/6$, the logarithmically enhanced density of states instead produces superconducting correlations consistent with an instability at arbitrarily weak attraction, although the accessible system sizes do not allow us to establish $|U_c|=0$ directly. Thus, the location and nature of the pairing instability are strongly controlled by the noninteracting band geometry, while the resulting ordered state remains a uniform onsite $s$-wave superfluid.

Over the interaction ranges investigated, we find no evidence of long-range charge order at any of the three fillings. At $\rho=2/3$, this conclusion is further supported by the strong-coupling description. In this regime, the fermion pairs map onto hard-core bosons at density $\rho_b=1/3$, with their hopping and nearest-neighbor repulsion constrained to satisfy $V_b/t_b=2$. This physical ratio lies far inside the superfluid regime of the bosonic model and well below the superfluid-to-valence-bond-solid transition near $(V_b/t_b)_c\simeq7.8$~\cite{Isakov2006}. Importantly, increasing $|U|/t$ reduces both $t_b$ and $V_b$ by the same factor and therefore does not drive the effective model toward the density-ordered regime.

These findings place a stringent constraint on the attractive Hubbard model as a minimal description of kagome metals such as AV$_3$Sb$_5$. A purely local and instantaneous attraction as in the case of the attractive Hubbard model can account for a uniform spin-singlet pairing tendency, but it does not produce the robust charge order found in these materials. If the attractive interactions are retarded and mediated by phonons instead, one arrives at the kagome Holstein model. There, at low phonon frequencies, the model exhibits a $K$-point charge-ordered phase at $\rho=2/3$~\cite{Bradley2023}. In the antiadiabatic limit, however, its retarded interaction becomes an instantaneous onsite attraction, and the model approaches the attractive Hubbard limit studied here. The Holstein CDW region therefore cannot persist to arbitrarily large phonon frequency, since its existence relies essentially on finite phonon dynamics and retardation.

Moreover, the $K$-point $\sqrt{3}\times\sqrt{3}$ charge pattern reported for the local Holstein model~\cite{Bradley2023} differs from the in-plane $M$-point $2\times2$ ordering observed in $A$V$_3$Sb$_5$~\cite{Jiang2023}. Taken together, these results indicate that reproducing the intertwined charge and superconducting phenomenology of the kagome metals requires ingredients beyond a single-band onsite attraction, such as nonlocal electronic interactions, momentum- or bond-dependent electron-phonon couplings, multiorbital structure, or lattice distortions. We leave these generalizations for a future study.

\acknowledgments
Computations by R.M. used resources from the Research Computing Data Core at the University of Houston and the TAMU ACES at Texas A\&M HPRC through allocation PHY240046 from the Advanced Cyberinfrastructure Coordination Ecosystem: Services \& Support (ACCESS) program, which is supported by U.S. National Science Foundation grants 2138259, 2138286, 2138307, 2137603, and 2138296. R.M.~acknowledges that the research was funded in part by the Robert A.~Welch Foundation, Grant \#L-E-0001-19921203.

\section*{Data availability}
The data that support the findings of this article are openly available in Zenodo~\cite{MondainiZenodo2026}.

\appendix

\section{Kinetic energy and double occupancy}
\label{sec:loc_proxy}
\begin{figure}[!tb]
  \includegraphics[width=1\columnwidth]{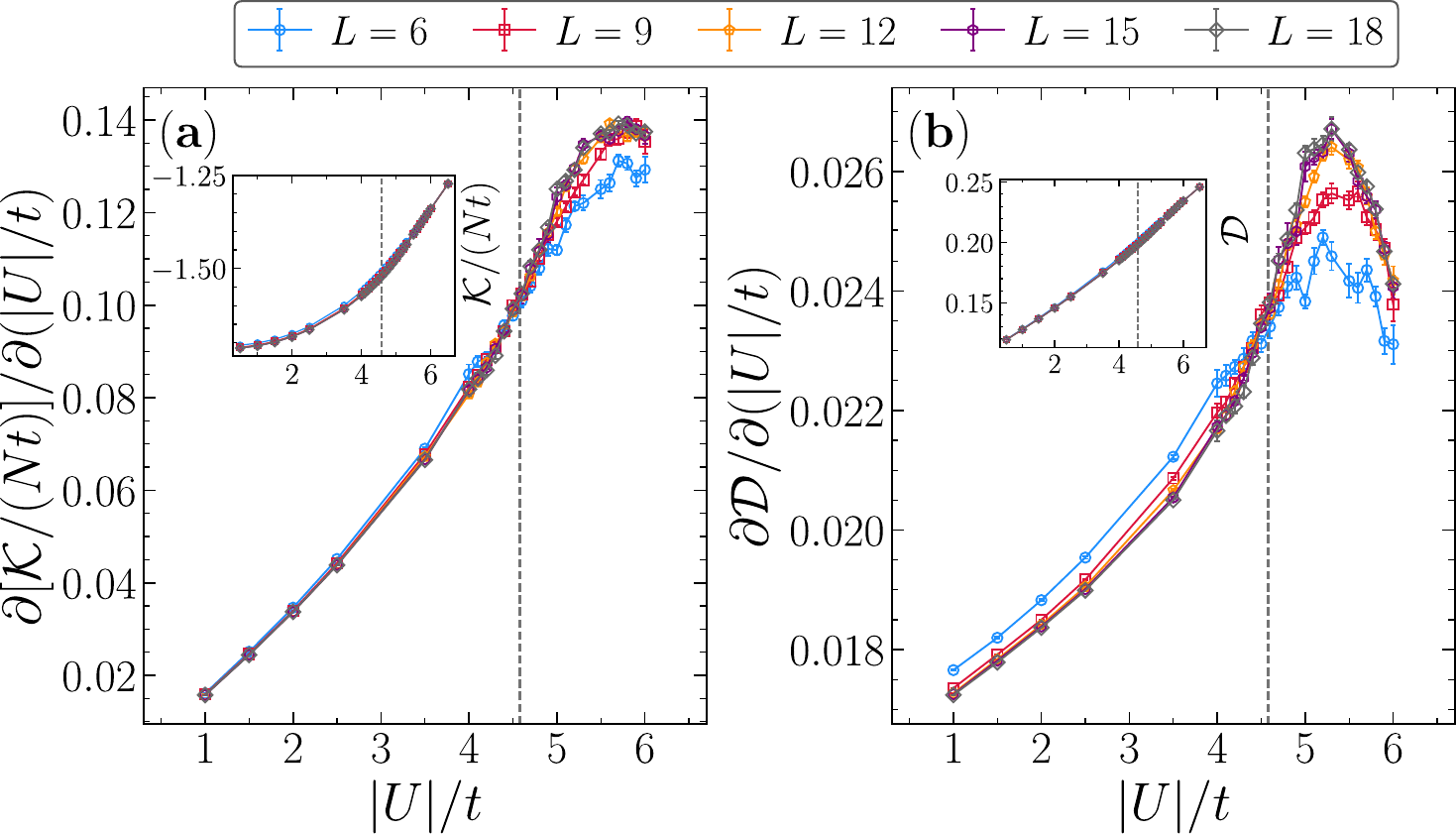}
\caption{Interaction dependence of the kinetic energy and double occupancy at $\rho=2/3$. (a) Derivative of the kinetic energy per site, $\mathcal{K}/(Nt)$, and (b) derivative of the double occupancy, $\mathcal{D}$, with respect to $|U|/t$ for the indicated lattice sizes. The insets show the corresponding undifferentiated observables. The gray dashed line marks the critical coupling $|U_c|/t=4.58$ obtained from the finite-size scaling analysis in Fig.~\ref{fig:scaling-ps}.}
  \label{fig:appendix_kenergy}
\end{figure}

We complement the order-parameter and spectral analyses with two local
observables: the kinetic energy $\mathcal{K}$ and the double occupancy,
\begin{equation}
    \mathcal{D}
    =
    \frac{1}{N}
    \sum_i
    \left\langle
    \hat n_{i\uparrow}\hat n_{i\downarrow}
    \right\rangle .
\end{equation}
Although neither quantity is an order parameter or provides an independent determination of the critical coupling, their interaction and system-size dependence offer a useful local characterization of the redistribution of correlations across the semimetal-to-superfluid transition. Figure~\ref{fig:appendix_kenergy} therefore shows $\partial[\mathcal{K}/(Nt)]/\partial(|U|/t)$ and $\partial\mathcal{D}/\partial(|U|/t)$, while the corresponding undifferentiated quantities are shown in the insets.

As the attraction increases, the double occupancy grows smoothly, reflecting the progressive formation of onsite pairs. More revealing is the change in finite-size dependence of $\partial\mathcal{D}/\partial(|U|/t)$  across the critical region. On the semimetallic side, $|U|<|U_c|$, this response decreases with increasing system size, whereas on the superfluid side, $|U|>|U_c|$, it increases with system size. The finite-size flow of this local response thus reverses in the vicinity of the independently determined critical coupling. Similarly, the kinetic-energy derivative exhibits little size dependence in the semimetallic regime but develops a clearer system-size dependence upon entering the superfluid phase.

These changes provide useful local trackers of the reorganization of the energetics and onsite correlations across the transition. We do not, however, use the locations of their broad extrema to estimate $|U_c|$, which is instead obtained from the finite-size scaling of the superfluid order parameter in Fig.~\ref{fig:scaling-ps}.

\bibliography{ref_kagome}

\end{document}